# AI-Guided Quantum Material Simulator for Education. Case Example: The Neuromorphic Materials Calculator 2025

Santiago D. Barrionuevo*[1,2] (ORCID: 0000-0002-4870-343X)

Myriam H. Aguirre*[2,3,4] (ORCID: 0000-0002-1296-4793)

[1] Instituto de Investigaciones Fisicoquímicas, Teóricas y Aplicadas (INIFTA), Universidad Nacional de La Plata – CONICET, Sucursal 4, Casilla de Correo 16 (1900), La Plata, Argentina

[2] Instituto de Nanociencia y Materiales de Aragón (INMA), CSIC–Universidad de Zaragoza, C/ Pedro Cerbuna 12, 50009, Zaragoza, Spain

[3] Laboratorio de Microscopías Avanzadas, Universidad de Zaragoza, Mariano Esquillor s/n, 50018, Zaragoza, Spain

[4] Departamento de Física de la Materia Condensada, Universidad de Zaragoza, C/ Pedro Cerbuna 12, 50009, Zaragoza, Spain

*Corresponding Authors : santi.barri@unizar.es , maguirre@unizar.es

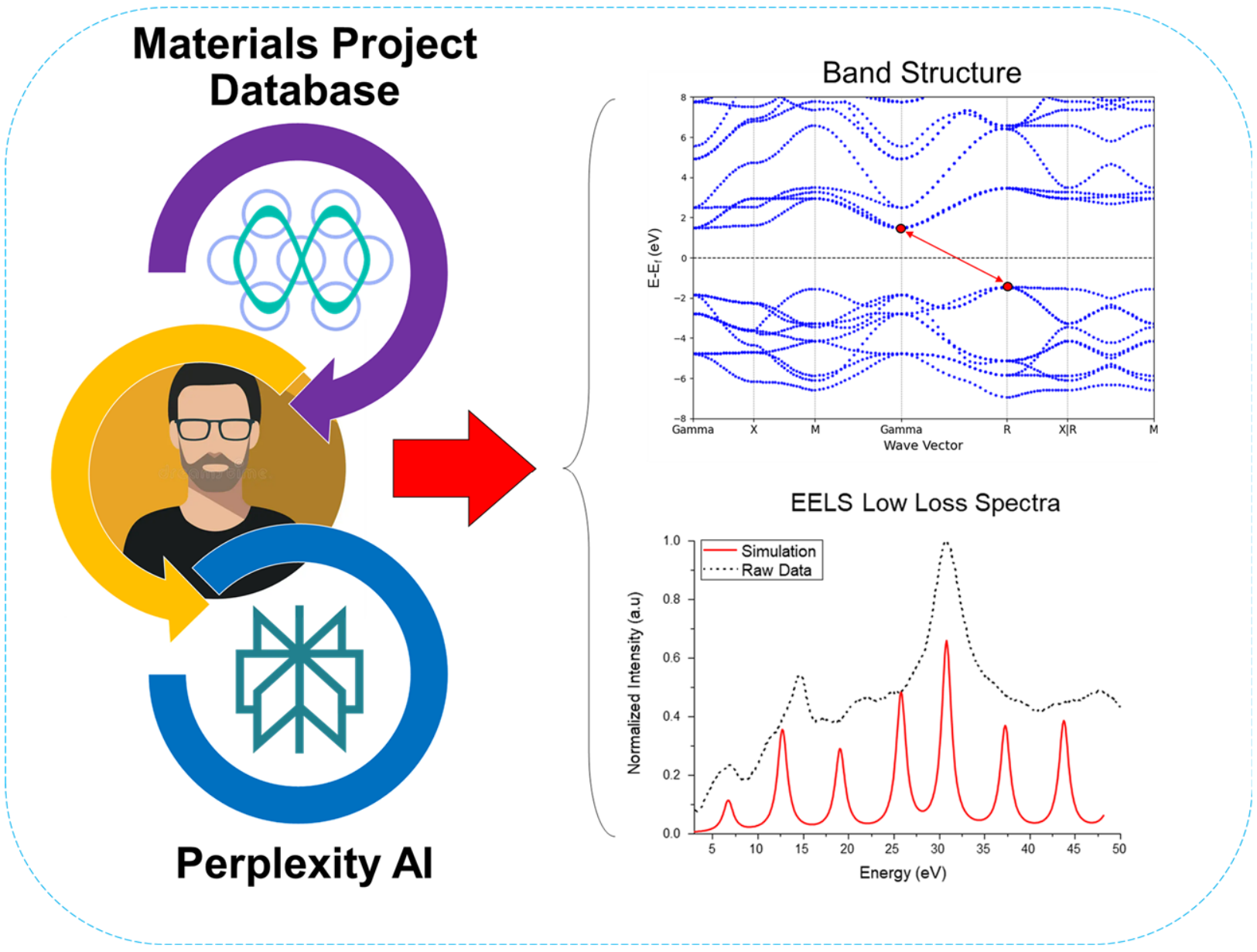

**TOC. Neuromorphic Materials Calculator 2025**

**ABSTRACT.** *Teaching and learning in advanced materials science are often limited by two barriers: the technical complexity of quantum-mechanical simulations and the lack of individualized support in inquiry-based education. Here, we introduce the* ***Neuromorphic Materials Calculator 2025 (NMC2025)****, a command-line platform that integrates a conversational artificial intelligence (AI) tutor with automated simulation workflows. NMC2025 combines large language model (LLM) guidance, real-time literature feedback, and domain-specific computation to create an adaptive learning environment. The system includes modular Python components for material discovery, simulation parameter optimization, and automated input generation for Quantum ESPRESSO (QE). Grounded in constructivist pedagogy, the tool enables students to carry out authentic research tasks such as identifying candidate materials for neuromorphic memristors or tuning density functional theory (DFT) inputs, while receiving context-aware explanations from the AI tutor. A case study illustrates how iterative, AI-guided refinement of hypotheses and calculations enhances both accuracy and understanding. NMC2025 fosters deeper conceptual insight, independent exploration, and smooth transfer of research methods into the classroom. This approach highlights the potential of AI-augmented education to reduce barriers to complex simulations and to expand access to computational modeling across science, technology, engineering, and mathematics (STEM).*

## INTRODUCTION

Modern science and engineering education increasingly emphasizes authentic learning experiences, where students learn by engaging in the same practices as researchers.(Herrington & Oliver, 2000) In materials science, this often means performing quantum-mechanical simulations (e.g., Density Functional Theory (DFT) calculations) to investigate properties of novel materials. However, mastering such simulations poses high cognitive load for students due to complex theoretical concepts, steep learning curves for software like Quantum ESPRESSO(Giannozzi et al., 2009), and the need to interpret results critically.(Sweller et al., 1998) Simultaneously, providing individualized guidance in these advanced topics is challenging in typical classroom settings.(Brown et al., 1989) Artificial intelligence (AI) offers a promising means to bridge this gap by acting as a virtual tutor or assistant that can guide students through complex tasks.(Kulik & Fletcher, 2016) Recent advancements in AI (e.g., large language model chatbots) enable tools that can answer questions, provide on-demand explanations, adapt to learner input, and even draw from up-to-date literature to support learning.(Zawacki-Richter et al., 2019) These AI-powered virtual assistants have been shown to facilitate personalized learning experiences and skill development for students, while reducing instructors' load.(D'mello & Graesser, 2013) At the same time, the educational value of simulation-based learning in science has been well established.(De Jong & Van Joolingen, 1998) Constructivist learning theories hold that students develop deep understanding by actively performing relevant tasks with appropriate scaffolding and feedback.(Jonassen, 1991) In the context of science education, computer simulations provide a powerful medium for such engagement, allowing students to explore "what-if" scenarios and visualize abstract phenomena.(Wieman et al., 2008) When coupled with guided inquiry, simulations can yield conceptual learning gains equal or superior to traditional instruction.(Jonassen, 1991) A review of 95 studies found that incorporating simulations into inquiry-based activities effectively supports learners' conceptual understanding of science topics, especially when combined with timely scaffolding.(Rutten et al., 2012) The challenge, however, is providing that scaffolding and expert feedback at scale. This is where an AI tutor can fill a crucial role, by giving students immediate, context-specific guidance as they conduct simulation experiments.

The Neuromorphic Materials Calculator 2025 (NMC2025) was developed to realize this synergy of AI guidance and simulation-based learning for an emerging domain: neuromorphic computing materials. Neuromorphic computing – brain-inspired computation – relies on novel materials like memristors (resistive memory devices) to emulate synaptic behavior.(Marković et al., 2020) These memristive materials exhibit continuously tunable resistance states, making them candidates for "neurosynaptic components" for building low-power, brain-like chips.(Jo et al., 2010) Training students to contribute to this cutting-edge field requires bridging fundamental theory (e.g., solid-state physics, DFT) with practical skills in materials modeling. NMC2025 addresses this need by combining an AI conversational agent (contextualized in materials science) with automated first-principles simulation pipelines. The main goal is to enable knowledge transfer from research to the classroom – students can tackle realistic research questions in a safe, guided environment, with the AI as a mentor and the simulation engine as a sandbox for experimentation. Centering our work in exploring and identify promising neuromorphic materials using AI-driven literature search and materials databases. Iteratively refining simulation parameters with AI-provided feedback and citations. Automatically generating and running Quantum

ESPRESSO input files to obtain results (e.g., electronic band structures, densities of states, electron energy loss spectra).

This work presents the **Neuromorphic Materials Calculator 2025 (NMC2025)**, a novel intelligent learning environment aligned with AI models, educational design, and curricular integration. We detail the system design and each Python module's role in supporting learning, situating the design within a pedagogical framework of inquiry-based, constructivist learning.(Jonassen, 1991) We then describe illustrative case studies of usage, demonstrating how NMC2025 supports adaptive scaffolding and student modeling through the AI's interactions. Finally, we discuss broader implications for AI in STEM education and outline future extensions (e.g., to other domains and longitudinal deployment).

## METHODS

***System Architecture.*** The Neuromorphic Materials Calculator 2025 (NMC2025) is a modular, command-line software suite designed to support AI-assisted quantum simulations in educational contexts. It integrates Python-based automation for Density Functional Theory (DFT) calculations with a large language model (LLM) via the Perplexity API to offer real-time tutoring, parameter critique, and literature citation. The system is organized into discrete modules corresponding to key stages in the simulation workflow: material selection, parameter definition, input file generation, and output analysis.

***Software and Dependencies.*** All modules are implemented in Python 3.10+ and use the following open-source packages:

- pymatgen(Jain et al., 2013) for structure manipulation and Materials Project API access.

- ASE(Hjorth Larsen et al., 2017) and NumPy(Harris et al., 2020) for data handling and formatting.
- OpenAI-compatible LLM queried via the Perplexity API(*Perplexity AI 2025*) for AI-driven guidance.
- Quantum ESPRESSO (v6.8) is used as the backend for electronic structure calculations.
- MPRester API is used to access crystal structures, computed properties, and standard pseudopotentials from the Materials Project database.(Jain et al., 2013)

***Operational Modes***

- ***Existing Materials Mode:*** Users input a known Materials Project ID (e.g., mp-5229). The software retrieves structural and electronic properties via API calls and generates simulation input files using generate_qe_inputs.py. The AI tutor analyzes the generated SCF input file and returns parameter suggestions with citations from peer-reviewed literature.
- ***Exploring Options Mode:*** Users provide a functional description (e.g., "phase-change material for neuromorphic switching"), and the AI performs a literature-backed search to recommend candidate materials. Each suggestion includes key physical descriptors and references. The system uses MPRester to cross-validate entries and retrieve structural data for input file generation.
- ***Pushing Frontier Mode:*** Users define custom, unpublished material systems by manually inputting lattice constants, atomic positions, element types, and DFT parameters. The assist_unpublished_material.py module interacts with the user through guided prompts, while generate_qe_inputs_ad_hoc.py builds the input files. The AI tutor then performs a review of all parameters and proposes optimizations based on comparable literature.

***Input File Generation and Validation.*** Input files for Quantum ESPRESSO include: scf.in, nscf.in, bands.in, bands_pp.in, and dos.in. The system ensures consistency across simulations by calculating derived quantities (e.g., number of bands based on total electrons). The analyze_input_with_perplexity() function sends each input to the AI for critique and proposes best practices, including recommended k-point meshes, cutoff energies, and smearing parameters. These recommendations are supported with scientific references.

***Output Analysis and EELS Simulation.*** Standard outputs (band structure, density of states) are parsed and optionally analyzed with AI_perp.py, which explains results and compares them with literature. A dedicated module, generate_qe_eels_inputs.py, automates input creation for Electron Energy Loss Spectroscopy (EELS) calculations using turbo_eels.x. This module adapts the SCF input to be compatible with EELS simulations and supports studies on dielectric response and plasmonic behavior in neuromorphic materials.

***Pedagogical Design.*** The system design is grounded in constructivist and inquiry-based learning models. Each mode fosters active engagement by requiring learners to input, justify, and revise simulation parameters. AI guidance is framed as “scaffolding” and adapts dynamically based on student choices. Literature citations are embedded in responses to promote scholarly practices and scientific reasoning.

***Case Study Validation.*** To demonstrate functionality, case studies were tested for each operational mode. These examples reflect real-world educational and research scenarios and illustrate typical system usage, including parameter exploration, simulation setup, AI-guided refinement, and results interpretation.

## RESULTS

NMC2025 is organized into modular Python components that together create an interactive educational workflow (Figure 1). Each block corresponds to a phase of the student's learning journey and is designed with specific AI-based instructional supports. **Fig. 1** shows the schematic architecture of NMC2025, organized as a modular pipeline. The system begins with the Welcome module, which orients the student and allows the selection of one of three operational modes. In "Existing Material" mode, the system uses the ***generate_qe_inputs*** module to prepare inputs directly from a known material ID. In "Exploring" mode, the ***assist_known_material*** AI agent helps students identify candidate materials, providing either a Materials Project ID or suggesting a new material, and proceeds to input generation accordingly. In "Pushing Frontier" mode, the ***assist_unpublished_material*** agent assists the student in defining parameters for a novel material, saving optimized configurations for ad hoc input generation. At critical stages, an AI agent (a large language model accessed via the Perplexity API) is invoked to suggest materials, critique simulation parameters, and analyze input files. The generated input files are then processed by the Quantum ESPRESSO backend to perform the calculations, with the outputs available for student analysis, optionally assisted by the ***AI_perp*** module. Arrows indicate the flow of information and the iterative feedback loops guided by the AI to refine the simulations.

**Fig. 2** shows the Welcome Module (welcome.py) of the Neuromorphic Materials Calculator 2025 (NMC2025), which serves as an introductory narrative and text-based user interface. It presents a welcome screen that outlines the tool's objectives and introduces the core mission of simulating neuromorphic materials for brain-inspired computing. The screen explains foundational concepts in neuromorphic electronics, such as the role of memristors—resistive switching devices that emulate synaptic behavior—and highlights the limitations of traditional digital systems. It also introduces the program's three modes of operation: Existing Materials, Exploring Options, and Pushing Frontiers, together with its integration into a simulation workflow based on Quantum ESPRESSO.(Giannozzi et al., 2009) Although the AI tutor is not yet active at this stage, the module provides educational context, activates prior knowledge (e.g., memristor function), and defines the simulation objectives. After this screen, the student selects one of the three simulation modes to proceed.

These three modes are described in detail in the following section.

***Existing Materials.*** This mode enables students to simulate a well-characterized material by providing its Materials Project ID.(Jain et al., 2013) Upon receiving the ID, NMC2025 retrieves detailed structural and electronic information from the Materials Project database and invokes the AI assistant to evaluate the key simulation parameters necessary for a Density Functional Theory (DFT) calculation using Quantum ESPRESSO.(Giannozzi et al., 2009) The system then automatically generates all requisite input files (e.g., for SCF, NSCF, band structure, and density of states calculations). Students are presented with the proposed parameters, along with AI-generated justifications grounded in the scientific literature, and are invited to either accept the configuration or iteratively refine it. This mode represents the most streamlined workflow in NMC2025, tailored for simulations of well-established materials commonly referenced in research or known to be suitable for neuromorphic applications.

***Exploring Options / AI-Guided Material Discovery.*** In this mode, the student can ask the program, without having specific materials in mind, to find candidates for a given application. For example, a prompt might be: “*Find materials for a memristor-based synapse in neuromorphic circuits.*” In this module, the student interacts with an AI assistant specialized in materials discovery. Internally, the script accepts a user’s description of the application or keywords and then formulates a query for the AI model (a large language model accessed via a Perplexity API client (*Perplexity AI 2025*)). The AI is instructed to prioritize neuromorphic-relevant materials and to present a numbered list of candidate materials with brief explanations, including key properties like chemical formula, band gap, and crystal structure. Crucially, the AI’s response is required to include full scientific references. This ensures the information is anchored in current literature and proper formatting, teaching students the importance of evidence-based reasoning (and exposing them to reading scientific references). The module uses the Materials Project database API (via MPRester) to cross-check any identified candidate by formula or ID. This means if the AI suggests a known compound (e.g., “$VO_2$ with metal-insulator transition”), NMC2025 will fetch its data from the authoritative database (band gap, structure, etc.). The AI’s suggestions are printed with color-coded formatting (e.g., cyan text for lines with DOIs) to enhance readability. After receiving suggestions, the student can either refine the search by providing additional criteria (the system prompts for more context if needed) or select one of the candidates for further investigation. Selecting a material trigger either retrieval of its Materials Project ID and data or, if it’s a novel suggestion without a database entry, the system falls back to an ad hoc path, in this mode it proposes parameters based on analogous materials, integrating AI-driven recommendations with expert user input. By stepping through this process, students practice inquiry skills: they learn how to translate an application need into materials criteria, how to interpret properties like band gap or crystal structure, and how to iteratively improve their query. The AI acts as a guide in a virtual library, helping students navigate both a vast search space and the nuances of materials data – essentially an intelligent librarian and domain expert combined. This component aligns with the concept of AI as a virtual agent guiding situated learning in simulations, as identified in prior literature.(Dai & Ke, 2022)

***Pushing Frontiers / Expert Simulation Setup Assistant.*** Aimed at advanced learners (e.g., graduate students) who want to simulate a new or "*unpublished*" material. Here, the student already has a material in mind (perhaps a hypothetical compound they synthesized) and needs to determine optimal simulation parameters. The AI assistant takes on the role of an expert research mentor in a dialogue designed to optimize a Quantum ESPRESSO (Giannozzi et al., 2009) DFT simulation . The script prompts the user for all essential inputs needed to define the system: material name, description, any known important properties, lattice parameter, plane-wave cutoff energy (ecutwfc), k-point meshes for self-consistent field (SCF) and non-SCF calculations, number of atomic species and their details (element symbol, atomic mass, pseudopotential file), total number of atoms, and their fractional coordinates in the unit cell. This mirrors the preparatory work a researcher would do when setting up a new simulation. As the student inputs these values, the system immediately computes derived suggestions – for instance, it calculates the recommended number of electronic bands (nbnd) based on electron count and whether the material is metallic or insulating. This real-time feedback is a form of embedded scaffolding within the tool, sparing students from manual formula calculations and alerting them to considerations (e.g., more bands needed for metals).(Jonassen, 1991) Once the initial parameter set is gathered, the module composes a detailed prompt to query the AI for evaluation. The prompt encapsulates the current simulation settings in natural language (*e.g., "Simulating crystal X with lattice = 4.2Å,*

*ecutwfc = 50 Ry, k-mesh 6×6×6, nbnd = 40, band gap = 0 eV, etc.*). The AI then returns recommendations and literature citations, which are displayed under a header "AI Expert Recommendations". For example, the AI might point out that the plane-wave cutoff is low for that class of materials and suggest a higher value, citing a relevant study (with DOI) that used a similar material. It might recommend refining the k-point density for better convergence, or note that if the material is magnetic, a spin-polarized calculation (nspin=2) should be used. All such advice comes with references, training the student to justify simulation choices with scientific sources. The student is then asked if they are satisfied or if they want to refine the parameters. If not satisfied, an iterative loop begins: the student can adjust any parameter on the fly (the program prompts for updates to lattice, cutoff, band gap, nbnd), and the AI re-evaluates the new parameters. This loop continues until the student accepts the recommendations. Throughout this process, adaptive scaffolding is in effect: the AI's feedback is tailored to the student's specific input and updates as the student makes changes, akin to a tutor responding to a student's revised solution attempt.(Jonassen, 1991) The module finally saves the converged "optimized" parameters to a JSON file for downstream use. By engaging with this expert assistant, students practice scientific decision-making: they see how small parameter choices can impact results and learn to consult literature for validation. The conversational format lowers the barrier to entry into research-grade simulation – instead of trial-and-error in isolation, the learner has a safety net of expert advice at each step. This aligns with educational best practices of providing timely and substantive feedback during complex tasks.(Jonassen, 1991) Moreover, requiring the student's active input (rather than automating everything) keeps them in control of the learning process, supporting a constructivist approach where the learner "constructs" the simulation setup with guidance.

**Simulation Input Generators.** Once a material and its parameters are determined via either of the above pathways, NMC2025 automates the creation of simulation input files. Two closely related modules handle this: "*generate_qe_inputs.py*" for known materials (with a database ID) and "*generate_qe_inputs_ad_hoc.py"* for ad hoc cases (with custom parameters). Both serve the function of translating the student's choices into concrete Quantum ESPRESSO(Giannozzi et al., 2009) input decks (text files for SCF, band structure, etc.), while also providing another opportunity for AI-driven instruction.

*Known Materials Input Generation (generate_qe_inputs.py):* This script is typically invoked after a student selects a candidate with a valid Materials Project ID (e.g. "mp-XXXX"). It fetches comprehensive material data via the Materials Project API (structure, formula, density, band gap, etc.). Using pymatgen, the structure is loaded and basic quantities like number of atoms, composition, and symmetry are derived. The script then computes a recommended number of electronic bands just as in the previous module, ensuring consistency. Next, an interactive loop begins where the student can adjust high-level parameters for the simulation: number of bands (with the recommendation as default), default k-point grids (SCF and NSCF), inclusion of Hubbard U for certain transition metal elements, and whether to enable spin-orbit coupling (if heavy elements are present). By prompting for these choices, the tool encourages students to think about *materials-specific physics* – for instance, recognizing if their material might require a Hubbard correction (e.g. if it contains iron or cobalt)(Wang et al., 2006) and learning the typical U values from defaults. After capturing user decisions, *generate_qe_inputs.py* automatically writes a set of input files: scf.in, nscf.in, bands.in (for band structure along high-symmetry k-path), bands_pp.in (post-processing for bands), and dos.in for density of states. All necessary blocks (SYSTEM,

ELECTRONS, ATOMIC_SPECIES, ATOMIC_POSITIONS, K_POINTS) are populated either from database values or student-specified parameters. The result is a ready-to-run simulation setup, achieved without the student having to manually format or calculate intricate input details. However, NMC2025 doesn't stop at generating these files; it leverages the integrated AI to review and explain the input files before execution. The function *analyze_input_with_perplexity()* reads the scf.in file and asks the AI to analyze it, effectively performing a *virtual code review* of the student's setup. The AI is prompted to compare the input against best practices from recent literature and to provide suggestions for improvement with references. This might result in feedback like: *"The k-point grid 6×6×6 may be too coarse for convergence; consider 8×8×8. Also, a smaller smearing (degauss) is recommended for insulators."* Such analysis reinforces learning by connecting the concrete input file to abstract principles (convergence, accuracy) and literature examples. The student can then choose to modify parameters (returning to the loop) or proceed to run the simulation. Notably, this create-review-adjust cycle embodies *learning by doing* with reflection: students make choices, see AI feedback, and have the agency to act on it, which is a powerful cycle for deep learning.

*Ad Hoc Input Generation (generate_qe_inputs_ad_hoc.py):* This module handles cases where the material is not in the database (likely following *assist_unpublished_material.py* or a non-database suggestion from *assist_known_material.py*). It reads the *optimized_simulation_parameters.json* produced earlier programs and then follows a similar process to build input files. One key difference is that here the lattice may not have full symmetry info; the script currently assumes a basic lattice (ibrav=1, cubic) unless additional info is given. Nevertheless, it performs validation and user engagement: it searches the local pseudopotential library for each element and, if multiple pseudopotential files exist (e.g. different exchange-correlation functionals), it asks the user to choose one. It also offers an interactive review of parameters (letting the user adjust lattice constant, cutoff, etc. before finalizing. This ensures that even in free-form cases, the student revisits their initial assumptions with a critical eye. After generating the same set of input files, this module can also incorporate a "literature evaluation" step– a placeholder in our current design where the AI could summarize how the chosen parameters align with or differ from known studies (this feature is noted for future enhancement). In essence, the ad hoc path mirrors the known-material path but relies more on user input and less on database defaults, appropriate for scenarios where the student truly explores the unknown.

**Extended Simulation Analysis (EELS Module).** In addition to standard SCF and band structure calculations, NMC2025 includes a specialized module *generate_qe_eels_inputs.py* that assists with setting up electron energy loss spectroscopy (EELS) simulations . Electron Energy-Loss Spectroscopy (EELS) probes a material's electronic structure by recording the energy lost by high-energy electrons that undergo inelastic Coulomb scattering as they traverse a thin specimen; analysis of the resulting loss spectrum yields elemental, chemical-state, and dielectric information at nanometer scales.(*Electron Energy-Loss Spectroscopy, EELS | Glossary | JEOL Ltd.*) This module modifies the SCF inputs (e.g. switching pseudopotentials if the chosen ones are incompatible with EELS calculations), then generates input files for low- and high-energy EELS spectra computations. By automating these steps, the tool frees students to focus on interpreting the resulting spectra rather than on tedious file preparations. In a classroom, students could use this to investigate how a material's dielectric response (captured in EELS) changes with doping or structure – again linking to authentic research questions.

**Evaluation via Case Studies.** To illustrate NMC2025's educational impact and the way it works, we present three simulated case studies based on plausible research and classroom scenarios. These case studies serve as a form of formative evaluation, demonstrating how the system might be used and what learning outcomes could be expected.

*Case Study 1 – Simulating Strontium Titanate ($SrTiO_3$) for Research Validation:* Alice, a second-year PhD student studying optical properties of strontium titanate ($SrTiO_3$) thin films, uses the Existing Materials operation mode of the Neuromorphic Materials Calculator 2025 (NMC2025) to simulate electronic properties and validate her experimental results. She initiates the simulation by entering the Materials Project ID mp-5229 for $SrTiO_3$. As described in **Fig. 3**, NMC2025 automates the generation of Quantum ESPRESSO input files, prompting Alice to confirm physical parameters such as Hubbard corrections and the number of bands. The AI tutor suggests applying a Hubbard U correction for Ti (U = 4.0 eV), a 6×6×6 k-point mesh for SCF, a 12×12×12 k-point mesh for NSCF, and an 80 Ry plane-wave cutoff, aligning with standard practices for $SrTiO_3$. It also calculates the required number of bands and confirms that spin-orbit coupling is unnecessary, as in this example . For a centrosymmetric, stoichiometric $SrTiO_3$ thin film whose optical gap, DOS and low-loss EELS are being benchmarked against room-temperature literature values, turning on SOC would slow the job dramatically but shift all key spectral features by only a few tens of meV—well below experimental resolution and the intrinsic DFT + U uncertainty. That is why the NMC2025 tutor correctly flags SOC as "unnecessary" for this particular research task. Alice reviews the AI's recommendations and generates input files for SCF, NSCF, DOS, bands, and EELS simulations using generate_qe_inputs.py. The system organizes these files into directories and confirms successful setup. To test the AI's suggestion, Alice runs simulations with and without the Hubbard U correction. She finds that the Hubbard correction improves the calculated bandgap and band structure, closely matching literature values. For EELS, she uses generate_qe_eels_inputs.py to create inputs for turbo_eels.x and simulates low-loss spectra to compare with her experimental data. NMC2025 plots key spectral features across different zones, enhancing Alice's understanding and enabling her to contrast her results with prior $SrTiO_3$ studies cited in ACS Nano style. The AI provides tabulated comparisons of SCF setups, k-point grids, pseudopotentials, and cell parameters, contextualizing each step with pedagogical guidance. This allows Alice to bridge theory and experiment, refine her computational skills, and interpret results confidently with AI-supported feedback based on results already published in the literature. This will not add new information, make suggestions, or conduct research on new materials for Alice. Rather, it serves as a support tool to assist the researcher in their scientific endeavors.

**Fig. 3** shows this Existing Materials operation mode of the Neuromorphic Materials Calculator 2025 (NMC2025), where the user (Alice) initiates a simulation by entering a Materials Project ID (e.g., mp-5229). This module automates the generation of Quantum ESPRESSO input files, prompting the user to confirm physical parameters such as Hubbard corrections and number of bands. Once confirmed, the system creates all necessary input files for SCF, NSCF, bands, DOS, and EELS simulations, organizing them into directories and confirming successful setup. Afterward, the AI assistant provides tailored guidance on best practices for first-principles simulations, offering parameter suggestions, literature references, and tabulated comparisons for SCF setups, k-point grids, pseudopotentials, and cell parameters. Although the AI is not yet running the simulation, it contextualizes each step with pedagogical scaffolding, helping learners understand how to properly configure their inputs. The user can revise parameters or proceed directly to launch the simulation, reinforcing the tool's dual purpose as both a research aid and an educational interface grounded in current materials science practice.

*Case Study 2 – Discovering a Neuromorphic Material:* Bob, a senior undergraduate researching memristors, uses the Exploration Mode of NMC2025 to identify materials with metal-insulator transitions suitable for neuromorphic switching. As shown in **Fig. 4**, this mode guides users through a structured selection process. Bob inputs his target property: "phase-change metal to insulator material for neuromorphic switching." The AI, leveraging its training and Materials Project data, returns a curated list of candidates, including (1) Vanadium(IV) oxide ($VO_2$) with a 68°C metal-insulator transition (band gap ~0.6 eV in the insulating phase) and (2) a chalcogenide glass with neuromorphic memory effects. Each candidate includes a brief description and a literature reference. Intrigued by $VO_2$, Bob selects it for simulation. NMC2025 retrieves its Materials Project ID, displaying its formula, 0.6 eV band gap, and rutile crystal structure. The AI suggests 60 bands for $VO_2$'s 24 electrons and notes its metallic phase (band gap < $1e^{-3}$ eV, with is_magnetic flag true), recommending spin-polarized calculations. Bob accepts most defaults but applies a Hubbard U correction for V atoms (U = 3 eV), as prompted for correlated oxides. The AI analyzes the generated scf.in file, citing a 2020 study recommending a 12×12×12 k-point mesh for $VO_2$. Bob, initially using a 6×6×6 mesh, updates to 12×12×12 and regenerates inputs. He runs the SCF and band structure calculations on a computing server. Using NMC2025's AI_perp.py, Bob interprets the output. The AI explains that the band structure shows a ~0.5 eV gap at the Fermi level, consistent with $VO_2$'s insulating phase but underestimated due to DFT limitations (experiments report ~0.6 eV). The U = 3 eV correction improved the gap size, as without it, the gap would be near zero. The AI provides a reference to support this analysis. Through this process, Bob identifies a promising material, learns the rationale behind simulation parameters, and gains insight into memristive behavior, supported by AI-guided exploration.

**Fig. 4** shows the Exploration Mode of the Neuromorphic Materials Calculator 2025 (NMC2025), where user (Bob) investigate and compare candidate materials for neuromorphic applications. Unlike direct simulation by ID, this mode guides users through a structured selection process powered by the AI engine. It prompts the user to define a target property or function—such as resistive switching, ferroelectricity, or phase-change behavior—and displays a curated list of suggested materials with key descriptors (e.g., band gaps, structure types, and prototype labels). Upon selection, the tool retrieves relevant materials data and guides the user through simulation parameter setup, highlighting choices like Hubbard U, pseudopotentials, and exchange-correlation functionals. Once a material is selected (e.g., $Sb_2Te_3$ for phase-change switching), the AI tutor delivers tailored recommendations and best practices based on current literature. It emphasizes reproducibility and robustness, providing rationale for parameter selection, typical pitfalls, and references. The module concludes with a parameter summary table, specific suggestions, and publication-ready references, reinforcing good computational practices. This design supports both guided discovery and independent research, helping users explore the frontier of material innovation through informed, reproducible simulations.

*Case Study 3 – Optimizing a New Material Simulation:* Charles, a first-year graduate student, explores a hypothetical layered oxide (e.g., W-doped $VO_2$) designed for ferroelectric and memristive properties, not yet in any database. He uses the Pushing the Frontier of Science mode of NMC2025, as depicted in **Fig. 5** A,B, to simulate this custom material. Charles manually inputs the structure, including lattice parameters (~4 Å), atomic positions (W, V, O), and pseudopotentials. The AI assistant evaluates the setup, providing expert guidance based on current literature. The AI estimates the system is insulating (band gap ~2 eV) and suggests 40 bands (occupied/2 + 5), correcting Charles' initial guess of 20 bands. It also recommends a smearing of 0.01 Ry for SCF convergence, referencing the Quantum ESPRESSO manual. Through iterative dialogue, Charles refines parameters, ensuring consistency with experimental data and reproducibility. Once satisfied, he generates input files via the ad hoc module and runs DFT simulations. The resulting density of states (DOS) shows a band gap of ~1.8 eV. Using AI_perp.py, Charles asks the AI to compare this with known materials. The AI notes that pure $VO_2$ has a smaller gap in its insulating phase (~0.6 eV) and suggests that W-doping may increase the gap, providing a plausible explanation for the 1.8 eV result. The AI delivers a parameter summary table, specific suggestions, and ACS Nano-style references, reinforcing computational best practices. Charles gains not only simulation results but also a deeper understanding of DFT parameter effects and the importance of literature validation. The AI mentor accelerates his learning, transforming a potentially lengthy process into a guided, efficient exploration of computational materials science.

**Fig. 5** A,B shows the Pushing the Frontier of Science mode in the Neuromorphic Materials Calculator 2025 (NMC2025), a workflow designed for expert users creating or refining

unpublished materials. In this example, the user inputs a custom structure—W-doped $VO_2$, relevant for neuromorphic phase switching—by manually entering lattice parameters, atomic positions, and pseudopotentials. Once defined, the AI assistant evaluates the setup and provides expert guidance on best practices based on current literature, including suggestions for lattice constants, plane-wave cutoffs, k-point grids, and the number of bands. Each parameter is assessed for consistency with experimental data and reproducibility in high-precision DFT calculations.

These case studies highlight qualitative outcomes: students using NMC2025 learned domain content (materials properties, DFT concepts), process skills (how to set up and validate simulations), and gained exposure to current research knowledge (via AI-provided references). The **immediacy of feedback** and the **iterative dialogue** are central to these outcomes, reflecting known benefits of AI tutors in maintaining engagement and facilitating deliberate practice. The next step will be to measure learning gains more formally (e.g. comparing classes using NMC2025 vs. traditional instruction on a similar project). One could assess improvements in students' ability to justify simulation choices or in their understanding of materials science concepts post-usage. Additionally, student attitudes towards working with AI in this context (whether it increased their confidence or interest in the subject) are important evaluation and are going to be conducted at scale in further studies.

## DISCUSSION

This software provides tools with several implications for AI in STEM education. Firstly, our work demonstrates the feasibility and value of blending state-of-the-art AI with domain-specific simulations to create rich learning experiences.(Alam, 2023) By leveraging an LLM (through the Perplexity API(*Perplexity AI 2025*)) that can access and cite scientific knowledge, we ensured the AI's guidance was not a "black box" hallucination bot but a trustworthy verifiable source of information.(Bender et al., 2021) This is critical in education – students must learn to trust but also verify information. The design decision to include citations for every substantive AI claim reinforces scholarly practice and can alleviate concerns about AI accuracy in academic settings.(Holstein et al., 2018) And most importantly, the role of the AI in NMC2025 is that of a facilitator and enhancer of human learning, not a replacer of the human or the learning or research task itself. The student remains central: making decisions, asking questions, and ultimately learning by doing. The AI does not do the assignment for them; it partners with them. This partnership model could be applied to many other STEM scenarios – from an AI-assisted chemistry lab (where an AI suggests synthesis steps or safety precautions) to a physics tutor that helps students set up and solve other complex simulations (e.g., finite element analysis with guidance analogous to what we present).(Woolf, 2009)

NMC2025's design is grounded in a constructivist and inquiry-based learning paradigm.(Jonassen, 1991) At its core, the tool positions the learner as an active participant—mirroring a researcher formulating questions, running simulations, and interpreting results—while the AI tutor and automation components provide scaffolding and help minimize cognitive load by handling repetitive or technically complex tasks.

This section analyzes how the system implements key principles of AI in education and how it can be integrated into curricula.

***Adaptive Scaffolding and Student Modeling.*** In educational theory, scaffolding refers to support given to learners that is gradually removed as they become more proficient.(Jonassen, 1991) NMC2025 provides scaffolding in multiple adaptive layers. The AI assistant modulates the difficulty by tailoring its feedback to the student's inputs – effectively modeling the student's current approach and needs. For instance, if a student chooses a very low cutoff energy (indicating a possible misunderstanding of accuracy requirements), the AI flags this with a warning and reference, essentially diagnosing a knowledge gap (perhaps the student didn't know typical cutoffs for that material class) and offering a remedy. Conversely, if a student already picks reasonable parameters, the AI might respond with confirmation and more subtle optimization tips, adjusting to a higher assumed knowledge level. This resembles an intelligent tutoring system (ITS) approach, where the AI's responses are akin to a cognitive tutor that "knows" the ideal solution and guides the student from their current state towards it.(Kulik & Fletcher, 2016) While NMC2025 does not explicitly maintain a user model over the long term (e.g., it does not store past student performance or misconceptions yet), it implicitly performs step-wise student modeling by analyzing each input and providing context-appropriate feedback. The inclusion of references in AI feedback also addresses metacognitive scaffolding – it encourages students to verify information and engage with external resources, building skills in self-directed learning beyond the immediate problem.(Azevedo & Hadwin, 2005)

***Inquiry-Based and Constructivist Learning.*** The tool is designed to facilitate inquiry cycles. In a typical use, a student might start with a broad question ("Which material could be good for…?" or "How do I simulate this new material?") and then proceed through hypothesis generation (AI suggesting candidates or parameter sets), experiment design (setting up simulations), observation (running calculations and obtaining results), and reflection (AI analyzing inputs or, potentially, outputs). NMC2025 supports the open-ended nature of inquiry by not enforcing a single correct path. Students can ask new questions mid-stream (refine search criteria, try a different material, etc.), and the AI will accommodate these deviations – much as a human tutor would when a learner's curiosity leads them slightly off-script. This flexibility is crucial for constructivist learning, where learners build their own understanding by exploring and connecting concepts.(Jonassen, 1991) The multimodal feedback (textual explanations with citations, data files, potential visualizations of output) caters to different learning styles and helps students make connections between the theoretical (textbook knowledge of DFT) and the practical (i.e seeing a band gap value emerge from a simulation they set up or the experimental measurement itself). Furthermore, by engaging with real research tools (Materials Project database(Jain et al., 2013), Quantum ESPRESSO(Giannozzi et al., 2009), scientific literature), students are constructing knowledge in a real-world context, which promotes transfer of learning(Lave & Wenger, 1991). This approach aligns with situated cognition principles – the idea that knowledge is best learned in the context of use – and indeed the mapping review by Dai and Ke (2022) noted the importance of virtual agents guiding learners in realistic simulation environments.(Dai & Ke, 2022)

***Motivation and Engagement.*** One often overlooked aspect of integrating AI in education is its impact on student motivation. NMC2025 attempts to keep learners engaged by giving them a sense of control and accomplishment in a complex task. The gamified welcome screen and the scenario of contributing to a futuristic neuromorphic project provide an intrinsic motivation boost (appealing to students' interest in cutting-edge tech).(Deterding et al., 2011)

As students' progress, the immediate feedback loops can create a flow experience, where the challenge is balanced with support: tasks are complex, but help is always at hand.(Csikszentmihalyi, 1990) The AI's ability to pull in up-to-date information and even phrasing like "I'm searching for the best candidates…" lends a sense that the student is collaborating with a knowledgeable partner, not just following a script. Prior studies on AI chatbots in education have found that students appreciate prompt assistance and the conversational format,(D'mello & Graesser, 2013) which can increase time-on-task and reduce frustration. By designing the AI responses to be encouraging and iterative ("Let's refine further…"), NMC2025 embodies a supportive tutor persona. This can be particularly impactful in advanced STEM learning, where students often feel overwhelmed; a non-judgmental AI assistant can encourage persistence.(VanLEHN, 2011) Additionally, the system naturally integrates the teaching of scientific communication skills – every time the AI presents a reference-laden recommendation, it models how to justify arguments with evidence, subtly enculturating students into the practices of scientific discourse.(*Laboratory Life | Princeton University Press*, 1986)

***Curricular Integration.*** The NMC2025 tool can be incorporated into graduate or senior-undergraduate curricula in materials science, physics, or chemistry. For example, in a computational materials science course, an assignment could be framed around NMC2025: students might be tasked with using the tool to identify a new candidate material for a memristive device and then simulate its electronic properties. The instructor can use the JSON and input outputs to verify that students went through the process, but the real assessment can focus on reflection – e.g., a report where students discuss what they learned from the AI feedback and how they chose their final parameters. Because NMC2025 handles the heavy lifting of simulations, instructors can allocate more class time to conceptual discussions (like why a certain material has a larger band gap, or what the significance of a predicted memristor behavior is) rather than debugging input files. Moreover, the tool's modular nature means it can support different learning objectives: one lab session might emphasize literature research skills (using assist_known_material to perform a mini literature review on candidate materials, guided by the AI's citations), whereas another session might emphasize technical skills in DFT (using assist_unpublished_material to understand how various simulation settings affect outcomes). The ability to adapt the tool's usage to different aims makes it a flexible addition to the curriculum. Importantly, using NMC2025 in class also addresses the often-cited gap between classroom science and actual scientific practice.(Chinn & Malhotra, 2002) Students get to see and use the same databases and software that researchers use. This authentic exposure can inspire students and give them confidence to transition into research roles.

***Multimodal Learning.*** While NMC2025 currently operates in a text-based terminal environment, it inherently produces multiple forms of output: written explanations, data files, and also visualizations (plots of DOS, BANDS, EELS, etc.). Integrating these into a more unified interface could further enhance learning. For example, a future version might have a GUI where the AI's advice is shown side-by-side with a live plot of the band structure as it's computed, with the AI pointing out features ("See that band crossing at the Fermi level? That indicates metallic behavior."). Research on multimodal tutors suggests that such synchronized visual and verbal feedback can cater to a wider range of learners and help in concretizing abstract concepts.(Mayer, 2009) Thus, one opportunity is to extend NMC2025 with visualization dashboards – perhaps using Jupyter notebooks or a web interface – while maintaining the robust backend logic of AI + simulation. This would transform it from a command-line tool into a more accessible learning platform.

Although NMC2025 is specialized for materials science, the underlying concept is broadly applicable. Any educational domain that involves complex workflows and expert decision-making could benefit from a similar marriage of AI tutoring and automation. For instance, in systems biology education, one could imagine a “Bioinformatic Calculator” where the AI guides students in setting up bioinformatics analyses or lab experiments (suggesting gene targets, experimental parameters, etc.), and then helps interpret the results.(Pevzner & Shamir, 2009) In engineering, an AI-driven CAD tutor might help students design and simulate circuits or structures, weaving in real-time advice based on engineering standards and prior designs.(Ullman, 2010) The key innovation is creating a continuous loop between AI guidance and hands-on activity, which keeps the student both informed and engaged. By experiencing this loop, students may also develop a healthier understanding of AI – seeing it not as an oracle, but as a tool that complements human creativity and expertise. Also , this addresses some concerns educators may have about AI (e.g., that educators will be replaced entirely by AI systems).(Chan & Tsi, 2023) In our design, passivity is not an option neither for students nor for teachers because the AI often asks the student to make a choice (it doesn’t autonomously decide to proceed without input). The student must reflect and respond and the teachers can also engage in this thinking game , thus everyone in the classroom stay cognitively active.

We also note that the integration of database resources (Materials Project in our case) shows how AI tutors can facilitate knowledge transfer between educational and professional contexts. Students using NMC2025 inadvertently become familiar with the Materials Project

Database, a real research tool, which could later empower them to use it independently in research. The success of this approach in materials science suggests similar integrations (e.g., an AI tutor integrated with arXiv or literature databases for a scientific writing class, guiding students in literature review).(Bornmann & Mutz, 2015) It points towards an ecosystem approach where AI educational tools are bridges linking learners with the vast resources available in the digital knowledge sphere. This also aligns with recent calls in AI-in-education research to leverage AI for enabling learner access to authentic scientific data and practices.(Black & Tomlinson, 2025; Picasso et al., 2024; Strielkowski et al., 2025)

## CONCLUSIONS

In this work, we introduce the Neuromorphic Materials Calculator 2025 (NMC2025), an innovative educational tool that demonstrates how AI tutors can be effectively embedded in real-world scientific computing tasks to deliver engaging and impactful learning experiences. By emphasizing neuromorphic materials and quantum simulations, NMC2025 makes an advanced research field accessible, turning it into an interactive learning platform through automated workflows and AI-driven guidance. Its modular design—with specialized components for each phase of exploration and adaptive AI assistance—provides a flexible framework for similar tools in advanced STEM education. This work encompasses not only the software and a practical case study but also a comprehensive teaching model that combines LLMs with scientific databases to foster inquiry-driven curricula in materials science. As AI technologies advance, we anticipate the rise of more human-AI collaborative learning environments crafted with educational purpose, enhancing rather than supplanting human intellect. NMC2025 embodies this vision, positioning AI as a mentor in the cycle of scientific discovery learning. By integrating AI tutoring, quantum simulation, and diverse feedback mechanisms, it establishes a foundation for next-generation educational tools that equip students for future roles in science and engineering. Importantly, it also cultivates the ability to critically assess AI-generated information, a vital skill in the age of AI-supported knowledge work. NMC2025 thus marks a pivotal advancement toward the classroom of tomorrow, where cutting-edge research and intelligent support are seamlessly integrated into the learning process.

## Statements & Declarations

### Funding

We acknowledge the financial support of the European Commission through Marie Skłodowska-Curie Actions H2020 RISE with the projects MELON (Grant No. 872631) and ULTIMATE-I (Grant No. 101007825). SB also acknowledges financial support from CONICET through doctoral and postdoctoral fellowships.

### Competing Interest

The authors declare no conflict of interest.

### Author Contribution

All authors contributed to the conception and design of the study. SB programmed the code and carried out the simulations, data analysis, and drafted the manuscript. MA provided critical revisions, supervision, and project coordination. All authors read and approved the final manuscript.

### Ethics Approval

Not applicable.

### Consent to Participate

Not applicable.

### Consent for Publication

Not applicable.

## Data Availability

The datasets generated during and/or analyzed in the current study are available from the corresponding author upon reasonable request.

## Code Availability

The Neuromorphic Materials Calculator 2025 (NMC2025) source code is openly available at GitHub https://github.com/SantiagoBLP/NeuroCalc2025. Version 1.0.0, corresponding to this study, is archived on Zenodo at DOI [10.5281/zenodo.17143512]. The concept DOI [10.5281/zenodo.17143511] always resolves to the latest version.

## AI Use Disclosure

Large Language Models (LLMs) such as the Perplexity API were used as integral components of the NMC2025 system architecture to provide educational guidance and parameter recommendations within the software. They were not used for autonomous manuscript writing; final responsibility for all content rests with the human authors.

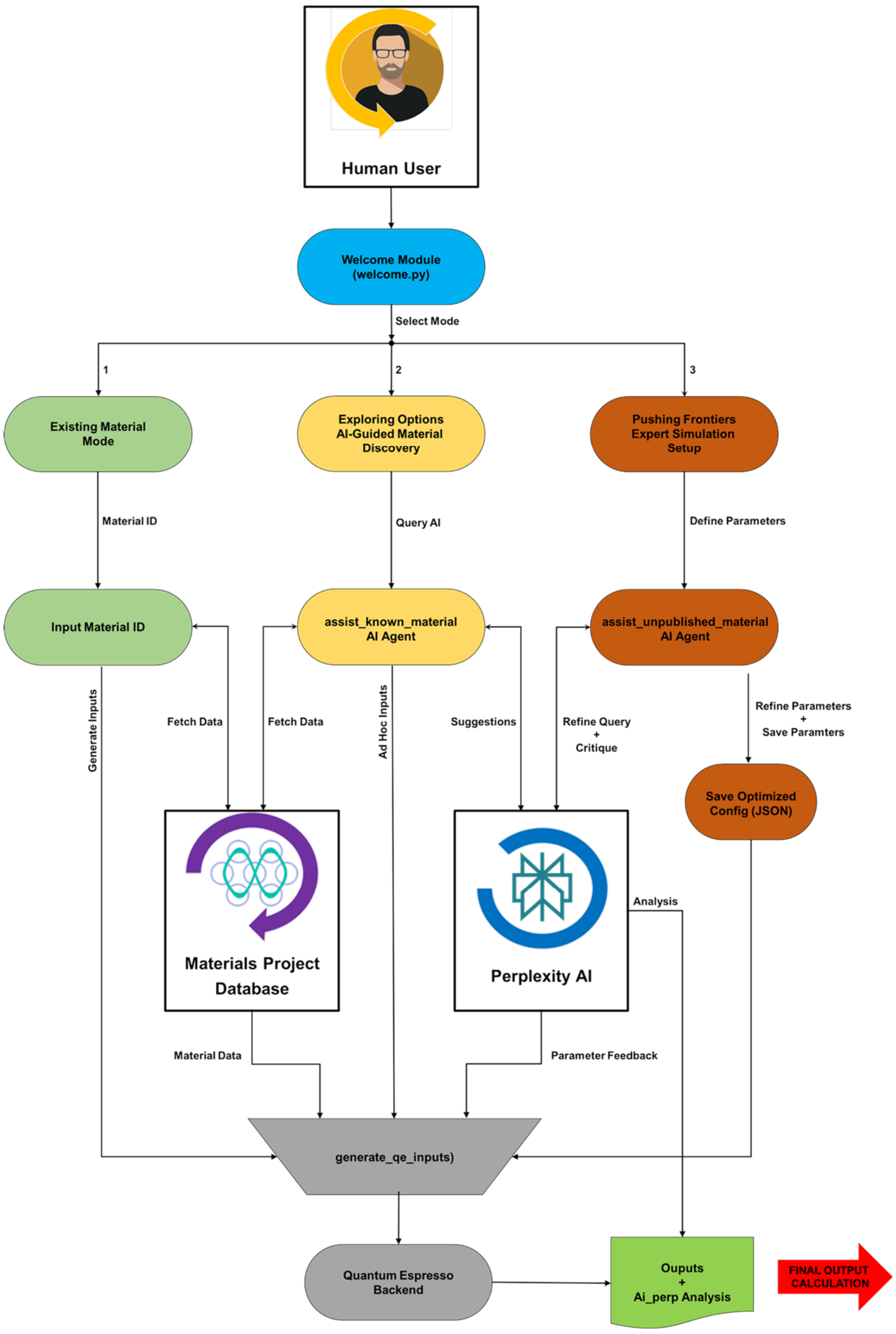

**Figure 1.** Schematic architecture of NMC2025. The system operates as a modular pipeline with three modes: Existing Material (direct input generation from a known ID), Exploring (AI-assisted candidate discovery and validation), and Pushing Frontier (guided setup for novel materials). At key stages, a large language model critiques parameters and suggests improvements. Quantum ESPRESSO executes the generated inputs, and outputs can be further analyzed with optional AI assistance. Arrows indicate information flow and iterative refinement.

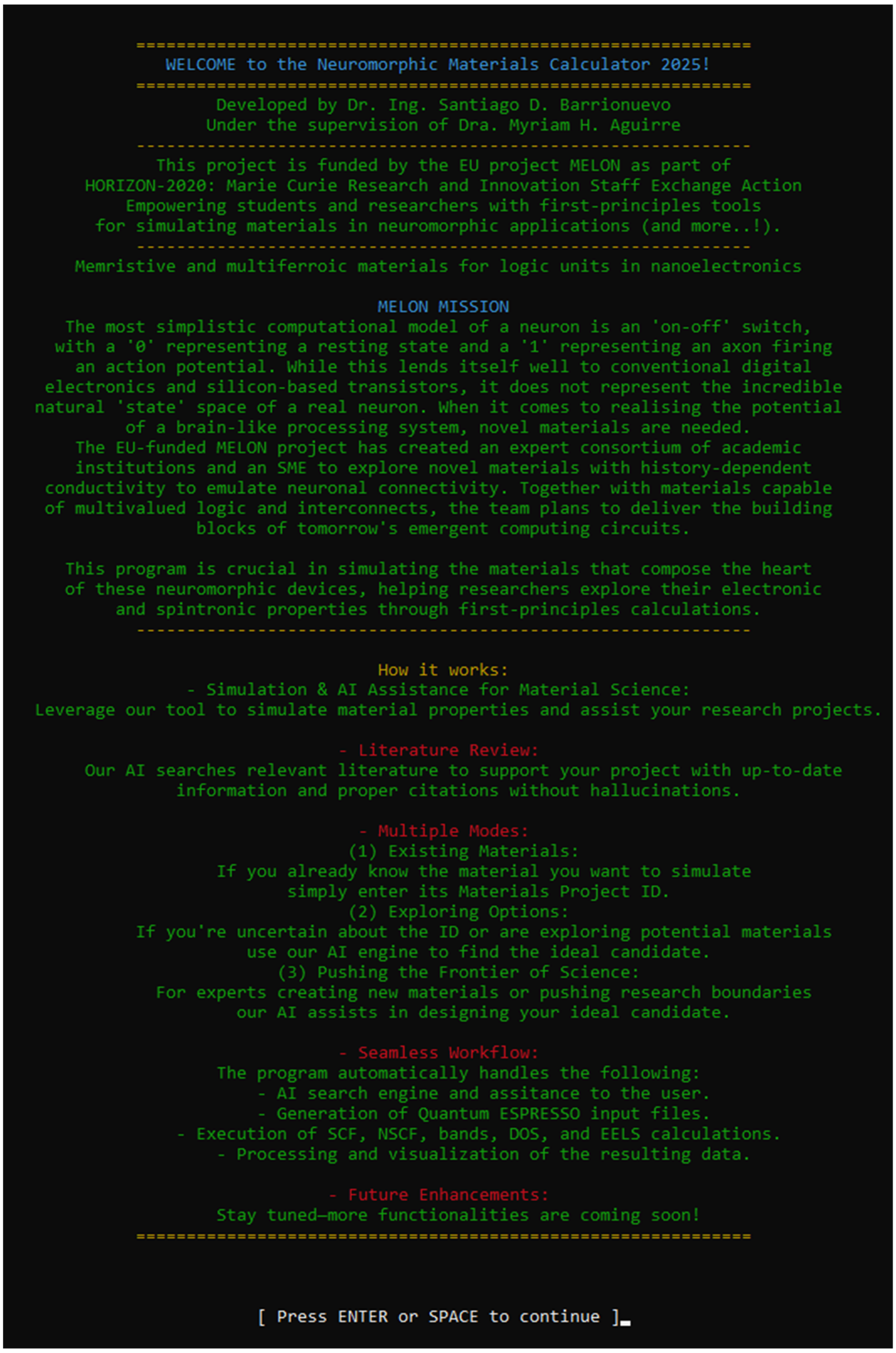

**Figure 2.** Welcome interface of NMC2025. The welcome.py module introduces the project context, outlines the three user modes (Existing Materials, Exploring Options, Pushing Frontier), and presents a text-based interface to engage learners in neuromorphic materials simulation.

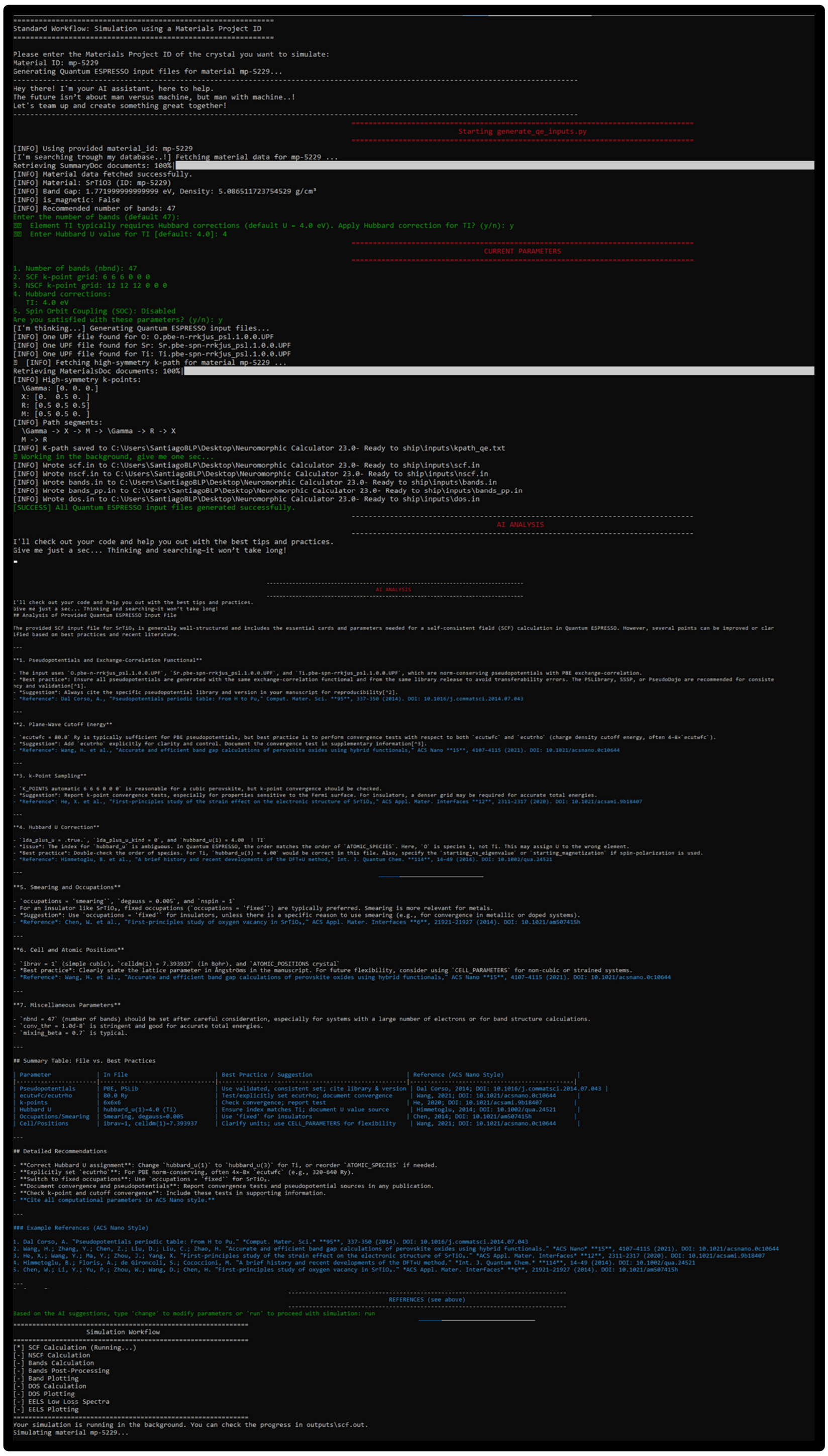

**Figure 3.** Simulation workflow for known materials using the "Existing Materials" mode in NMC2025. The figure illustrates the process of entering a Materials Project ID (e.g., mp-5229), generating Quantum ESPRESSO input files, and receiving AI-guided suggestions. The assistant reviews key simulation parameters (Hubbard U, number of bands, etc.), produces all required input files, and offers best practices and literature references to ensure high-quality, reproducible DFT simulations.

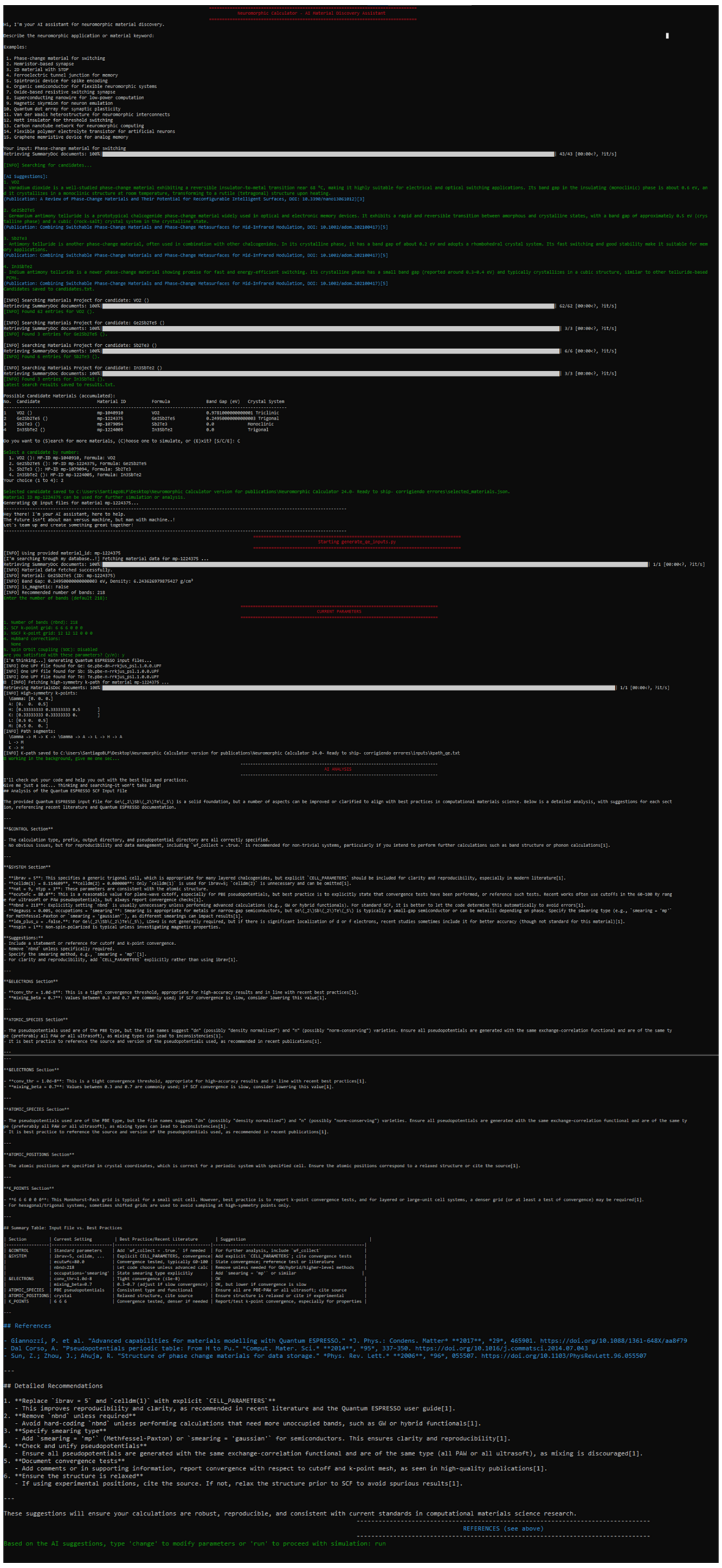

**Figure 4.** Exploration-based material discovery via the “Exploring Options” mode in NMC2025. Users begin by selecting a functional objective (e.g., resistive switching or phase-change behavior), prompting the AI to suggest candidate materials with relevant properties. Once a material is selected (e.g., $Sb_2Te_3$), the assistant provides simulation-ready parameters, rationale for each choice, fostering informed material selection for neuromorphic research.

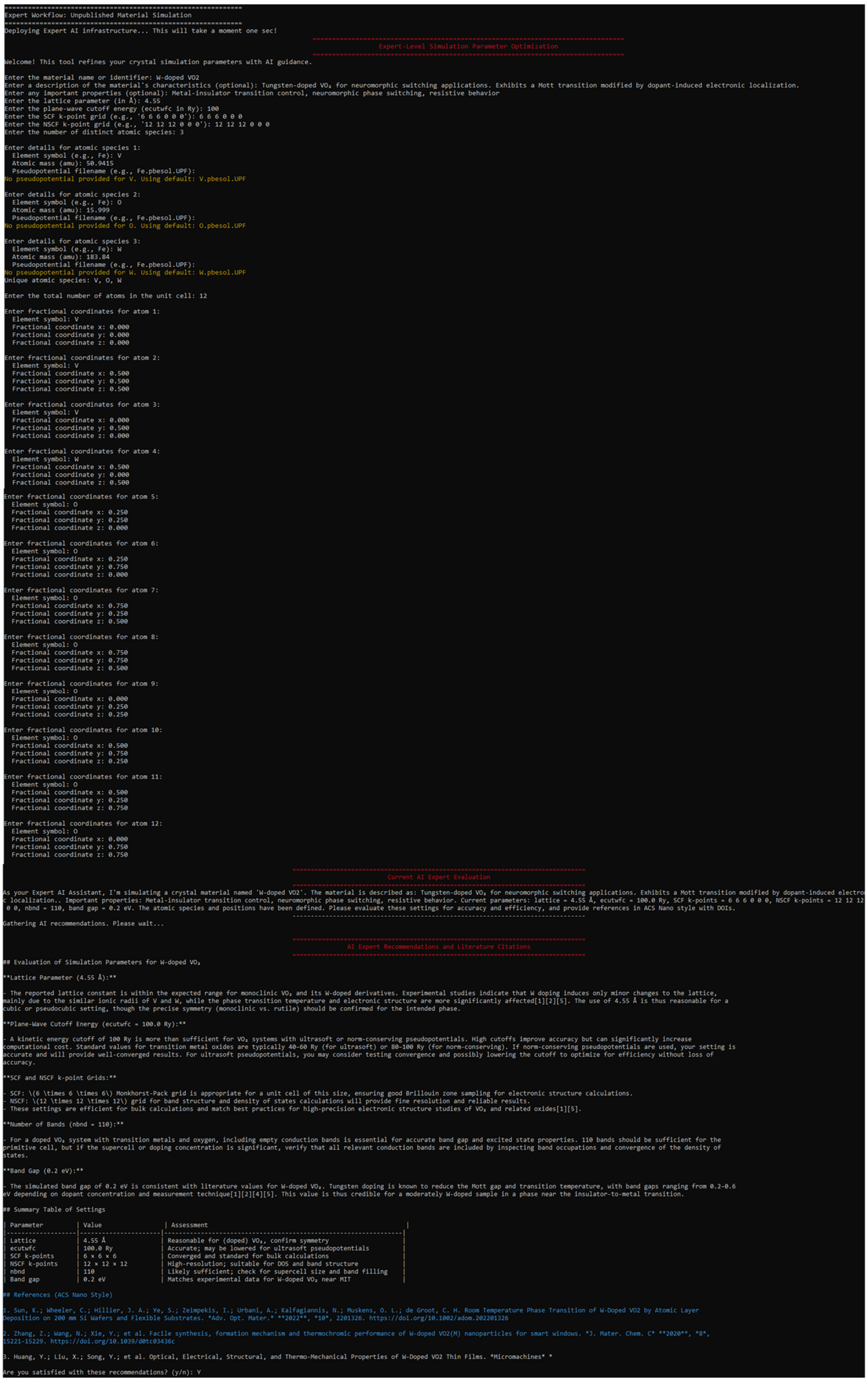

**Figure 5.** Expert simulation setup in the "Pushing the Frontier of Science" mode of NMC2025 (Part 1). Here, a user defines an unpublished or custom material—W-doped $VO_2$—by manually entering lattice constants, atomic positions, and pseudopotentials. The AI evaluates these inputs, prepares them for optimization, and begins validating parameters using domain-specific literature and simulation standards.

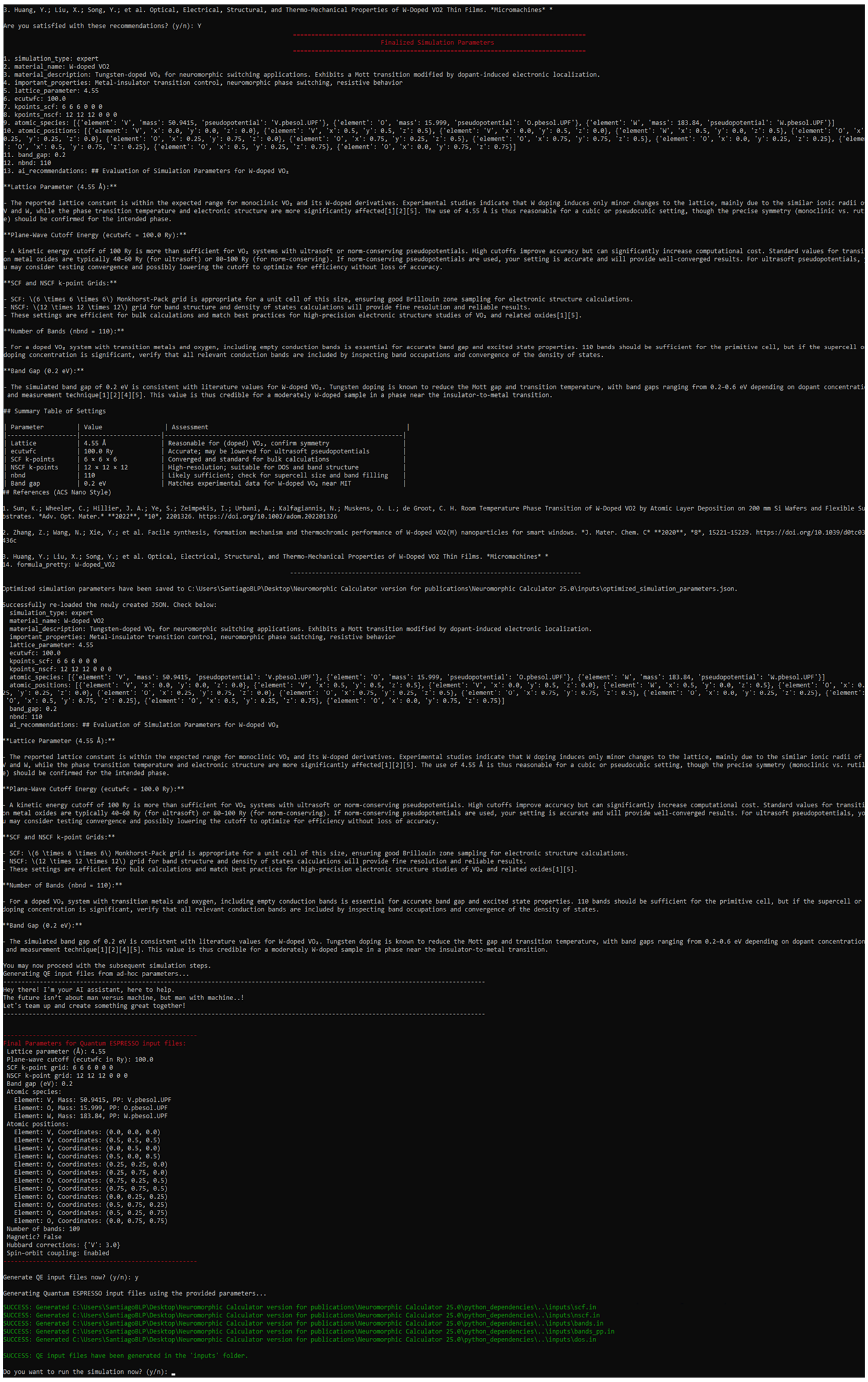

**Figure 5.** (B) Completion of the expert workflow for W-doped $VO_2$ in the "Pushing the Frontier" mode of NMC2025 (Part 2). The AI assistant reviews all input parameters, generates Quantum ESPRESSO files, and offers detailed assessments supported by cited experimental and computational studies. The user is given the option to proceed with the simulation, completing a process that mirrors expert-level materials modeling with an emphasis on reproducibility and scientific rigor.

## Glossary of Terms – NMC2025 Paper

| Term | Definition |
|---|---|
| **AI (Artificial Intelligence)** | Simulation of human intelligence by machines, enabling learning and decision-making. |
| **AI Tutor / AI Assistant** | A conversational agent using AI to guide students through tasks and provide scientific feedback. |
| **Band Structure** | Energy levels of electrons in a material, determining its conductive properties. |
| **Constructivist Learning** | Learning theory where knowledge is built through experience and reflection. |
| **Cutoff Energy (ecutwfc)** | Kinetic energy cutoff for plane-wave basis in DFT calculations. |
| **DFT (Density Functional Theory)** | Quantum method for calculating electronic structures of materials. |
| **DOS (Density of States)** | Number of electronic states at each energy level in a system. |
| **EELS** | Electron Energy Loss Spectroscopy used to probe electronic structure and bonding. |
| **Exchange-Correlation Functional** | Function used in DFT to approximate electron interactions. |
| **Fermi Level** | Energy level with 50% probability of electron occupancy at 0 K. |
| **Hubbard U** | DFT correction for better handling of electron correlations in certain materials. |
| **Input File** | Text file specifying simulation parameters for Quantum ESPRESSO. |
| **Intelligent Tutoring System (ITS)** | AI system offering personalized educational feedback. |
| **Inquiry-Based Learning** | Approach focused on student exploration and problem-solving. |
| **k-Point Mesh** | Grid used to sample reciprocal space in DFT. |
| **Kohn–Sham Equations** | Equations used in DFT to simplify electron interactions. |
| **Large Language Model (LLM)** | AI trained on massive text data to generate human-like language. |
| **Lattice Parameters** | Dimensions and angles defining a crystal unit cell. |
| **Materials Project** | Database with computed materials properties used in simulations. |
| **Memristor** | Memory device with resistance dependent on history; key in neuromorphic computing. |
| **Module** | A Python script handling specific tasks in NMC2025. |
| **NBND (Number of Bands)** | Number of calculated electronic bands in DFT. |
| **Neuromorphic Computing** | Brain-inspired computing using devices like memristors. |
| **NSCF** | Non-self-consistent DFT calculation for band structure or DOS. |
| **Perplexity API** | Interface connecting to an LLM for generating citations and suggestions. |
| **Pseudopotential** | Simplification in DFT replacing core electrons. |
| **Quantum ESPRESSO (QE)** | Software for electronic-structure calculations using DFT. |
| **Ry (Rydberg Unit)** | Energy unit used in DFT; 1 Ry ≈ 13.6 eV. |
| **SCF** | Self-consistent field calculation updating electron density in DFT. |
| **Smearing / degauss** | Technique to handle partial occupations near the Fermi level. |
| **Spin Polarization (nspin = 2)** | DFT setting to include magnetic effects. |
| **Text-Based Interface** | User interface based on command-line input/output. |
| **Unit Cell** | Smallest repeating unit in a crystal lattice. |